\documentclass[useAMS,usenatbib]{mn2e}

\usepackage{amsmath}
\usepackage{amssymb}
\usepackage[dvipsnames]{xcolor}
\usepackage{graphicx}
\usepackage{graphics}
\usepackage{hyperref}
\hypersetup{
     colorlinks   = true,
     citecolor    = blue
}

\def\apj{{ ApJ}}
\def\apjl{{ApJL}}

\def\aap{{ A\&A}}

\def\mnras{{ MNRAS}}
\def\araa{{ ARA\&A}}

\def\nat {{ Nature}}
\def\pasj{{ PASJ}}
\def\ssr{{ Space Sci. Rev.}}

\def\prd{{ Phys. Rev. D}}

\def\mt{\mathrm}
\def\flu{\mathcal{F}}
\def\md{\mt{d}}

\newcommand{\myemail}{wenbinlu@astro.as.utexas.edu}

\title[EDF for repeating FRBs]{A universal EDF for repeating
  fast radio bursts?}
\author[Lu \& Kumar]
  {Wenbin Lu$^1$\thanks{\myemail},
  Pawan Kumar$^1$\thanks{pk@astro.as.utexas.edu}\\
  $^1$Department of Astronomy, University of Texas at Austin, Austin,
TX 78712, USA}
\date{\today}

\pagerange{\pageref{firstpage}--\pageref{lastpage}} \pubyear{2016}
\def\LaTeX{L\kern-.36em\raise.3ex\hbox{a}\kern-.15em
    T\kern-.1667em\lower.7ex\hbox{E}\kern-.125emX}

\begin{document}
\label{firstpage}
\maketitle

\begin{abstract}
Assuming: fast radio bursts (FRBs) are produced by
neutron stars at cosmological distances; FRB
rate tracks the core-collapse supernova rate; and all FRBs repeat with a
universal energy distribution function (EDF) $\md 
\dot{N}/\md E \propto E^{-\beta}$ with a cutoff 
at burst energy $E_{\rm max}$. We find that
observations so far are consistent with a universal EDF with
$1.5 \lesssim \beta \lesssim 2.2$, high-end cutoff $E_{\rm
  max}/E_0 \gtrsim 30$ and normalization $\dot{N}_0 \lesssim 2\rm\
d^{-1}$; where $\dot{N}_0$ is the integrated rate above the
reference energy $E_0 \simeq 1.2\times 10^{39} f_{\rm r}^{-1} \rm\
erg$ ($f_{\rm r}$ is the radio emission efficiency). Implications
of such an EDF are discussed. 
\end{abstract}


\section{Introduction}\label{sec:intro}
Fast radio bursts (FRBs) are bright millisecond flashes mostly found at high
Galactic latitudes \citep{2007Sci...318..777L,
  2013Sci...341...53T}. Since discovery, they have received a large
amount of theoretical study on possible progenitors\footnote{These
  models include: collapsing neutron stars \citep{2014A&A...562A.137F,
    2014ApJ...780L..21Z}, mergers \citep{2013PASJ...65L..12T},
  magnetars \citep{2010vaoa.conf..129P, 
    2015ApJ...807..179P, 2015arXiv151204503K}, giant pulses from young pulsars
  \citep{2016MNRAS.458L..19C, 2016MNRAS.457..232C,
    2016arXiv160302891L}, shocks \citep{2014MNRAS.442L...9L}, flaring
  stars \citep{2014MNRAS.439L..46L}, asteroids colliding neutron
  stars \citep{2015ApJ...809...24G, 2016arXiv160308207D}. See
  \citet{2016arXiv160401799K} for a recent review.},  burst mechanisms
\citep[e.g.][]{2014PhRvD..89j3009K, 2016MNRAS.457..232C} and potential
usage to probe the intergalactic medium \citep[IGM, ][]{2014ApJ...780L..33M,
  2014ApJ...797...71Z} and cosmology \citep{2014ApJ...788..189G}. The
strongest argument for their cosmological 
origin is that the large dispersion measures (DMs), the line-of-sight
electron column 
density $DM = \int n_{\rm e} {\mt d}l \sim 10^3\rm\
pc\ cm^{-3}$, are inconsistent with being from the average interstellar
medium (ISM), stellar corona, HII regions and
supernova remnants (SNRs) in the Galaxy \citep[or even the Local
Group,][]{2014ApJ...797...70K, 2014ApJ...785L..26L}. The fact that FRB
121102 has been observed to 
repeat \citep{2016Natur.531..202S, 2016arXiv160308880S} rules out
catastrophic events at least for this burst.

FRB models are mostly based on neutron stars (NSs), which could
naturally accommodate the short durations $\Delta t\lesssim 1\rm\ ms$,
bright coherent emission, 
repetitivity on a range of intervals ($\sim10^2$ to $10^5\rm\ s$ or
longer, depending on the flux level). In addition, many FRBs
show pulse widths of $W \sim 1(\nu/\rm GHz)^{\sim-4}\rm\ ms$
consistent with multi-path propagation spreading and are much
longer than what intergalactic or Galactic scattering could account
for \citep{2013ApJ...776..125M,
  2016ApJ...818...19K}. This means the local plasma
surrounding the progenitors must be more strongly scattering than
the average ISM. The rotation measure given by the linearly polarized FRB
110523 is much larger than intergalactic or Galactic contributions,
meaning the progenitor is located in a dense magnetized
nebula \citep{2015Natur.528..523M}. The DM from a SNR decreases with
time, but the repeating FRB 121102 has a constant $DM\simeq 559\rm\ 
pc\ cm^{-3}$ for 3 yrs. This means a possible NS must be older than
$\sim 100\rm\ yr$ and the DM from the SNR is smaller than $\sim
100\rm\ pc\ cm^{-3}$ \citep{2016arXiv160404909P}. Therefore, FRBs are
likely from NSs of 100 yr to 1 Myr old embedded in SNRs or star-forming
regions.

Due to insufficient monitoring time, the other (so-far)
``non-repeating'' FRBs could also be repeating.
% If so, accurate sky
% positions could be obtained in the future and we will pin down their
% host galaxies.
In this {\it Letter}, we assume that FRBs are
from NSs at cosmological distances and that they repeat with a
universal energy distribution function\footnote{By ``universal'' we 
  mean the probability distributions of the free parameters are well peaked.} 
    (EDF). Below, we first summarize the properties of FRB 121102 in
    section \ref{sec:121102}, and then explore the answers to the
    following questions: (I) {\it Do FRB statistics so far support a
    universal EDF? If so, what constraints can we put on it?}
(II)  {\it Is FRB 121102 representative of the ensemble?} (III) {\it What is
    the spacial density of FRB progenitors?} 
\section{FRB 121102}\label{sec:121102}
The isotropic equivalent energy of each burst is
\begin{equation}
  \label{eq:4}
  E \approx
1.2\times10^{39} \frac{\flu}{\rm Jy.ms} f_{\rm r}^{-1} D_{\rm Gpc}^2 \Delta
\nu_9\rm \ erg, 
\end{equation}
where $\flu$ is the fluence, $f_{\rm r}$ is the radio emission
efficiency, $D_{\rm   Gpc} = D\rm\ Gpc^{-1}$ is the luminosity
distance\footnote{We use
  $\Lambda$CDM cosmology with $H_0 = 71 \ \rm km \ s^{-1} \
  Mpc^{-1}$, $\Omega_{\rm m} = 0.27$, and $\Omega_{\rm \Lambda} =
  0.73$.} and $\Delta \nu_9 = \Delta\nu \rm\ GHz^{-1}$ is the
bandwidth of the FRB spectrum. The DM from IGM is given by
\citep{2004MNRAS.348..999I}
\begin{equation}
  \label{eq:14}
 DM(z) = \int_0^z \frac{c\md z^\prime}{H_0\sqrt{\Omega_{\rm
       m}(1+z^\prime)^3 + \Omega_{\rm \Lambda}}} \frac{n_{\rm
     e}(z^\prime)}{(1+z^\prime)^2}
\end{equation}
and $n_{\rm e}(z) = 2.1\times10^{-7} (1+z)^3\rm cm^{-3}$ is the number
density of free electrons. FRB 121102 has
\begin{equation}
  \label{eq:5}
  \begin{split}
     DM &= DM_{\rm tot}  - DM_{\rm Galaxy} - DM_{\rm host} \\
  &\simeq 559 - 188 - 100 = 271 \rm\ pc\ cm^{-3},
  \end{split}
\end{equation}
which corresponds to redshift $z_0\simeq 0.28$ and luminosity distance
$D(z_0) \simeq 1.4\rm\ Gpc$.
% If ignoring the host galaxy's
% contribution $DM_{\rm host}$, we have $DM_{\rm IGM}\simeq 371 \rm\ pc\
% cm^{-3}$ and $z_0\simeq 0.38$ and $D(z_0)\simeq 2.0\rm\ Gpc$. 
The repeating rate above fluence $\flu_0 = 0.5\rm\ Jy.ms$ at 1.4 GHz is
$\int_{\flu_0} ({\mt d} \dot{N}/{\mt d} \flu) {\mt d} \flu \simeq
2\rm\ d^{-1}$. The progenitor's time-averaged isotropic equivalent
luminosity is
\begin{equation}
  \label{eq:1}
\dot{E}\simeq 6.5\times
10^{34} f_{\rm r}^{-1} D_{\rm Gpc}^2 \Delta \nu_9\rm \ erg\
s^{-1}\simeq 10^{35} f_{\rm r}^{-1}\rm\ erg\ s^{-1},
\end{equation}
if we add up the total burst fluence of $3.1 \rm\ Jy.ms$ during
Arecibo\footnote{The source is not simultaneously 
    monitored by different telescopes, but Green Bank Telescope
    (S-band) gives a similar result $\dot{E}\simeq 3.2\times
  10^{34} f_{\rm r}^{-1} D_{\rm Gpc}^2 \Delta \nu_9\rm \ erg\
  s^{-1}$.} Telescopes' on-source time $15.8\rm\ hr$. The energy
reservoir required to supply the bursting activity for a time $\tau =
10^3\tau_3 \rm\ d$ is 
\begin{equation}
  \label{eq:3}
  E_{\rm tot} \simeq 10^{43} f_{\rm r}^{-1} \tau_3 \Omega_{\rm
    r}/4\pi \rm \ erg. 
\end{equation}
where $\Omega_{\rm r}$ is the total sky coverage of all bursts from
one progenitor. If FRBs are concentrated only in a narrow
range of stellar latitude (e.g. poles), we have $\Omega_{\rm r}\ll 4\pi$.
The fluence distribution function of FRB 121102 is a
power-law ${\rm d} N/\mt{d} \flu\propto \flu^{-1.78 
  \pm 0.16}$ \citep{2016arXiv160408676W}, so $\dot{E}$ should mainly
come from the most energetic bursts (which we missed due to
insufficient monitoring time), 
and this is why eq.(\ref{eq:3}) is a lower limit. 
If the energy reservoir is the magnetosphere of a NS, it requires a
magnetic field strength of $B\gtrsim 1.6 \times 10^{13} (\tau_3\Omega_{\rm
    r}/4\pi f_{\rm r})^{1/2}\rm\ G$.

\section{A universal luminosity function}\label{sec:rate}
In this section, we assume FRB rate tracks the core-collapse supernova
rate and that all FRBs repeat with a universal EDF. We calculate the
detection rate of FRBs on the Earth as a 
function of source redshift, and then by 
comparing it with observations, we constrain the parameters of the EDF.

\subsection{Model}
Core-collapse supernova rate tracks the cosmic star-formation rate and
is given by \citep{2014ARA&A..52..415M}
\begin{equation}
  \label{eq:8}
  \Phi_{\rm cc}(z) = 2.8\times10^2 \frac{(1+z)^{2.7}}{1 + \left[(1+z)/5.9
\right]^{5.6}} \rm\ Gpc^{-3}\ d^{-1}
\end{equation}
We assume that a fraction $f_{\rm frb}$ of NSs are able to produce
observable FRBs and stay in the active phase for a local-frame time
$\tau$. During the active phase, a NS undergoes multiple ($\gtrsim 
10^3$) bursts  
intermittently and we call it a ``bNS'', short for ``bursting neutron
star''. The EDF is assumed to be a power-law with a high-end
cutoff\footnote{Since bursts 10 times dimmer than the reference
  fluence $\flu_0$ have been observed from FRB 121102, a possible
  low-end cutoff could be at $E_{\rm min}/E_0< 10$. As can be seen
  later from eq.(\ref{eq:11}), such a low $E_{\rm
  min}$ will be observationally noticed only for very nearby sources at distances
$D<0.31\mt{Gpc} \frac{D(z_0)}{1.4\rm Gpc}\sqrt{\left(10\frac{E_{\rm min}}{E_0}\right)
  \left(2\frac{\flu_0}{\flu_{\rm th}} \right)}$, which corresponds to
a $DM\lesssim 62\rm\ pc\ cm^{-3}$. Current observations have found no
FRBs below $200\rm\ pc\ cm^{-3}$ and hence put no constraint on the
possible low-end cutoff.}
\begin{equation}
  \label{eq:7}
  \frac{\mt{d}\dot{N}}{\mt{d}E} =
  \begin{cases}
    \frac{(\beta -1)\dot{N}_0/E_0}{1 - \left(E_{\rm
              max}/E_0\right)^{1-\beta}} \left(
          \frac{E}{E_0}\right)^{-\beta},\ &\mt{if}\ E < 
    E_{\rm max},\\
    0,\ &\mt{if}\ E > E_{\rm max},\\
  \end{cases}
\end{equation}
where $E_0$ is the reference burst energy corresponding to a fluence $\flu_0$
when the source is at redshift $z_0$ and $\dot{N}_0$ is the integrated
rate above $E_0$. Without losing generality, we base the reference
point ($E_0$, $z_0$ and $\flu_0$) on FRB 121102
\begin{equation}
  \label{eq:9}
  \ z_0 \simeq 0.28,\ \flu_0 = 0.5\
  \mt{Jy.ms},\ E_0\simeq 1.2\times10^{39}f_{\rm r}^{-1}\rm\ erg.
\end{equation}
where $z_0$ has an uncertainty of $\sim 30\%$ (due to unknown
$DM_{\rm host}$), the uncertainty of $E_0$ mostly comes from $z_0$
(through eq. \ref{eq:4}), and $f_{\rm r}$ is the radio emission
efficiency. The statistics of FRB 121102 give
$\dot{N}_0\equiv \int_{E_0}^\infty 
(\mt{d}\dot{N}/\mt{d}E) \md E \simeq 2\ \mt{d^{-1}}$, but we keep the 
normalization constant $\dot{N}_0$ as a free parameter, since
whether FRB 121102 is representative is to be determined. The other two
free parameters ($\beta$, $\xi\equiv E_{\rm max}/E_0$) are restricted
in the ranges $\beta > 1$ and $\xi > 1$, because weaker bursts are
more frequent and we have detected bursts with $E> E_0$.

Integrating over the cosmic volume, we get the all-sky detection rate
above a fluence threshold $\flu_{\rm th}$
\begin{equation}
  \label{eq:10}
  \dot{N}_{\rm det}(\flu_{\rm th}) = f_{\rm frb} \tau \int_0^{z_{\rm
      max}} \md z \frac{\Phi_{\rm 
      cc}}{1+z} \frac{\mt{d}V}{\mt{d}z} \int_{E_{\rm
      th}(z)}^{E_{\rm max}} \frac{\md \dot{N}}{\md E} \md E,
\end{equation}
where $\md V/\md z$ is the differential comoving volume, $E_{\rm
  th}(z)$ is the threshold energy above which bursts from a given
redshift $z$ are detectable
\begin{equation}
  \label{eq:11}
  \frac{E_{\rm th}(z)}{E_0} = \frac{\flu_{\rm th} D(z)^2}{\flu_0 D(z_0)^2},
\end{equation}
and $z_{\rm max}$ is the given by $E_{\rm th}(z_{\rm max}) = E_{\rm
  max}$. Combining eq.(\ref{eq:8})--(\ref{eq:11}), we calculate the
differential detection rate $\md
\dot{N}_{\rm det}/\md z$ as a function of redshift. We assume that the
dimensionless quantity $f_{\rm  frb}\tau\dot{N}_0$ 
does not depend on redshift, so the normalized cumulative distribution of
the all-sky event rate only depends on the two free
parameters ($\beta$, $\xi$). We use $\chi^2$
formalism to fit it with the observational normalized cumulative
distribution of DM, i.e.
\begin{equation}
  \label{eq:13}
  \left(\int_0^z \frac{\md \dot{N}_{\rm det}}{\md z^\prime} \md
  z^{\prime} \right)_{\rm n}
\mbox{v.s.}
\left( \int_0^{DM(z)} \frac{\md N_{\rm obs}}{\md DM^\prime} \md
  DM^\prime \right)_{\rm n}
\end{equation}
where $DM(z)$ is only the IGM component (eq. \ref{eq:14}) and we use a
constant host galaxy contribution\footnote{We have also tested the case
  where $DM_{\rm host} \equiv 0$ and the difference is small compared
  to the other uncertainties.} $DM_{\rm
  host} \equiv 100\rm\ pc\ cm^{-3}$. The allowed parameter space in
the ($\beta$, $\xi$) is determined by the significance probability
$P(\chi^2/\rm dof)$.

On the other hand, by matching the normalization at a
given $\beta$ and $\xi$ with the all-sky detection
rate from observations, we
calculate the product $f_{\rm frb}\tau \dot{N}_0$, so the allowed
parameter space in the $\beta$-$\xi$
plane constrains $f_{\rm frb}\tau \dot{N}_0$.

\subsection{Observations}\label{sec:obs}
In the FRB search strategy (de-dispersion $\rightarrow$
single-pulse search), the signal-to-noise ratio and hence
detectability depends on both the fluence and de-dispersed pulse width
as $S/N \propto \flu 
W^{-1/2}$ \citep{2015MNRAS.447.2852K}. If there is a
maximum intrinsic pulse width $W_{\rm max}$, the threshold fluence for
a given $(S/N)_0$ is $\flu_{\rm th} \propto (S/N)_0 W_{\rm
  max}^{1/2}$. \citet{2015MNRAS.447.2852K} derive a
completeness threshold of $2\rm\ Jy.ms$ for the Parkes
FRBs. Therefore, we test our model only on Parkes FRBs with $\flu >
\flu_{\rm th}= 2\rm\ Jy.ms$, which converts to a sample 
of 8 bursts \citep{2016arXiv160103547P}.

Some other factors could introduce biases: (I) scattering by the Galactic
ISM may bias against detection of FRBs at low Galactic
latitudes \citep{2014ApJ...792...19B, 2014ApJ...789L..26P}; (II) in
de-dispersion trials, typically, DMs of $\lesssim 100$ and $\gtrsim
2000 \rm\ pc\ cm^{-3}$ are not considered, resulting 
in effectively a low- ($<0.1$) and high-redshift ($>2.5$) blindness; (III)
the source position within a single beam ($\mt{FWHM} \simeq 15'$) is
unknown and most authors use the on-axis assumption and report a lower
limit. 

Due to (I), we have discarded FRBs that occurred below Galactic
latitude $20^{\rm o}$. It turns out (II) does not introduce
significant biases because the observed FRB are all between $0.25 <
z < 1.4$ and the true rate should cut off sharply below and above this
range. Based on (III), we also try the sample selected by $\flu_{\rm th} =
1\rm\ Jy.ms$ (with 11 FRBs) for comparison.

\begin{figure}
  \centering
\includegraphics[width = 0.4 \textwidth,
  height=0.37\textheight]{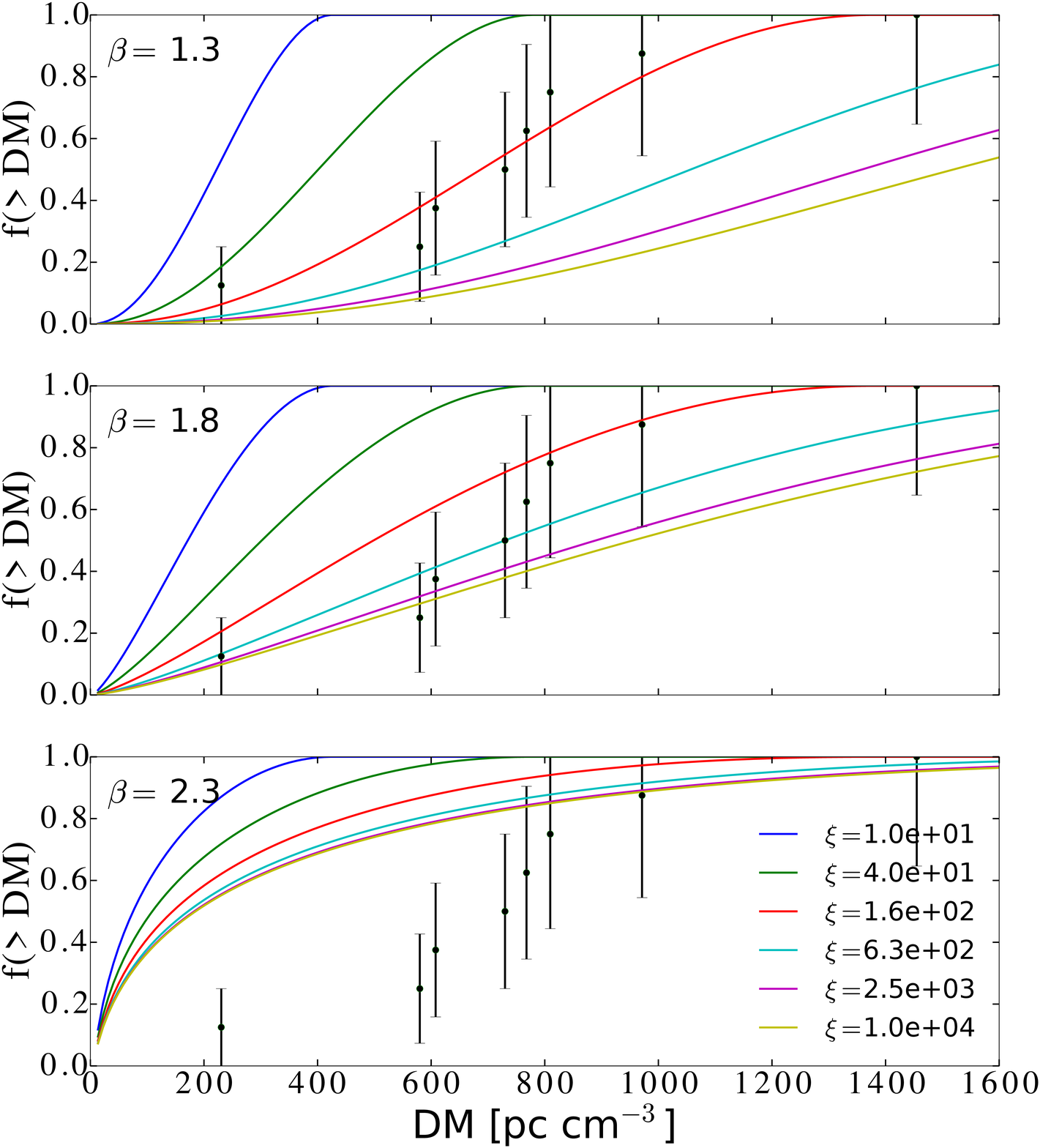}
\caption{Comparison between normalized theoretical and observational
  FRB rate distributions. A fluence threshold $\flu_{\rm th} =
  2\rm\ Jy.ms$ is used. 
% We fix the host galaxy's contribution
%   $DM_{\rm host} = 100\rm\ pc\ cm^{-3}$. 
}\label{fig:comp}
\end{figure}

\subsection{Results}\label{sec:result}
\begin{figure}
  \centering
\includegraphics[width = 0.45 \textwidth,
  height=0.35\textheight]{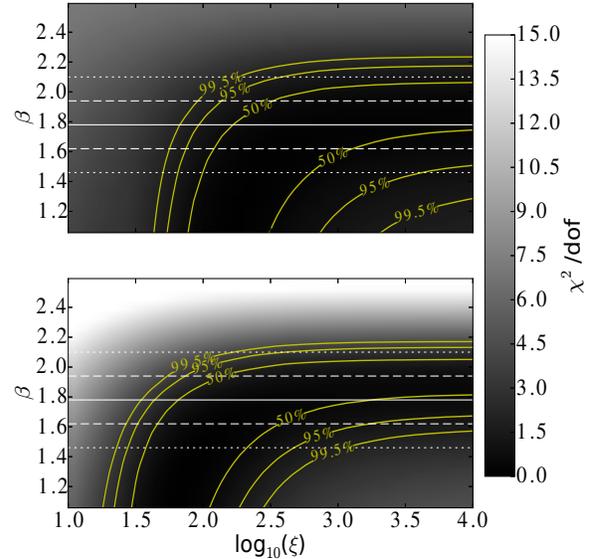}
\caption{The reduced $\chi^2$ of the fits between theoretical
  and observational FRB rate distributions. The upper and lower panels
  are for the two samples with $\flu_{\rm th} = 2$ and $1 \rm\ Jy.ms$
  respectively. The contours show the confidence levels. The power-law
  index for FRB 121102 $\beta = 1.78\pm 0.16$ is marked by a white solid
  line, with 1- and 2-$\sigma$ errors between dashed and dotted lines.
}\label{fig:chisq}
\end{figure}

In fig.(\ref{fig:comp}), we compare the theoretical event rate above
threshold $\flu_{\rm th} = 2\rm\ Jy.ms$ with observations, for three 
different power-law slopes ($\beta=1.3,\ 1.8,\ 2.3$) and various
energy cutoffs ($\xi\equiv E_{\rm max}/E_0$ from 10 to $10^4$). We can
see that, for a larger $\beta$ and smaller $\xi$, more FRBs are
expected to have small $DM$. 

We quantify the goodness of the fit by calculating the reduced
$\chi^2$, as shown in fig.(\ref{fig:chisq}). The upper panel is for
$\flu_{\rm th} = 2\rm\ Jy.ms$ (8 bursts) and the lower for $\flu_{\rm
  th} = 1\rm\ 
Jy.ms$ (11 bursts). We can see that the two samples generally agree
with each other and the allowed parameter space is
\begin{equation}
  \label{eq:16}
 \{ \beta \lesssim 2.2,\ \xi\gtrsim 10^{1.5} \}.
\end{equation}
The bottom-right corner with \{$\beta \lesssim 1.4$, $\xi \gtrsim
10^3$\} may be in tension with observations but the confidence level
is not high and more data is needed. We also mark the
power-law index for FRB 121102 $\beta = 
1.78\pm 0.16$, with 1- and 2-$\sigma$ errors between dashed and
dotted lines. The agreement with the constraints from the whole sample
means that FRBs may indeed share a universal EDF. FRB 121102 may be
representative in that its power-law index 
lies within the range allowed by the statistics from the whole
population. 

\begin{figure}
  \centering
\includegraphics[width = 0.4 \textwidth,
  height=0.17\textheight]{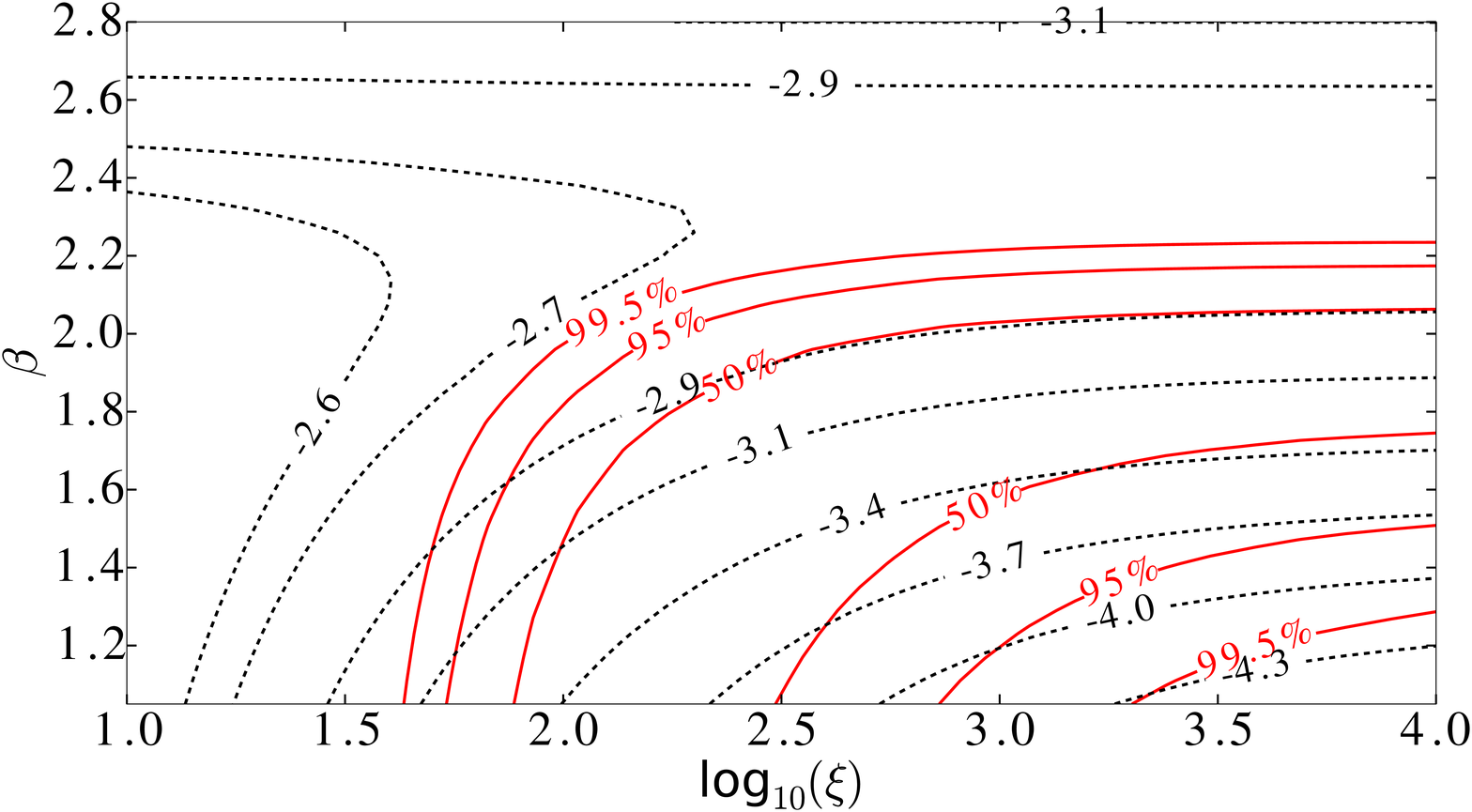}
\includegraphics[width = 0.4 \textwidth,
  height=0.17\textheight]{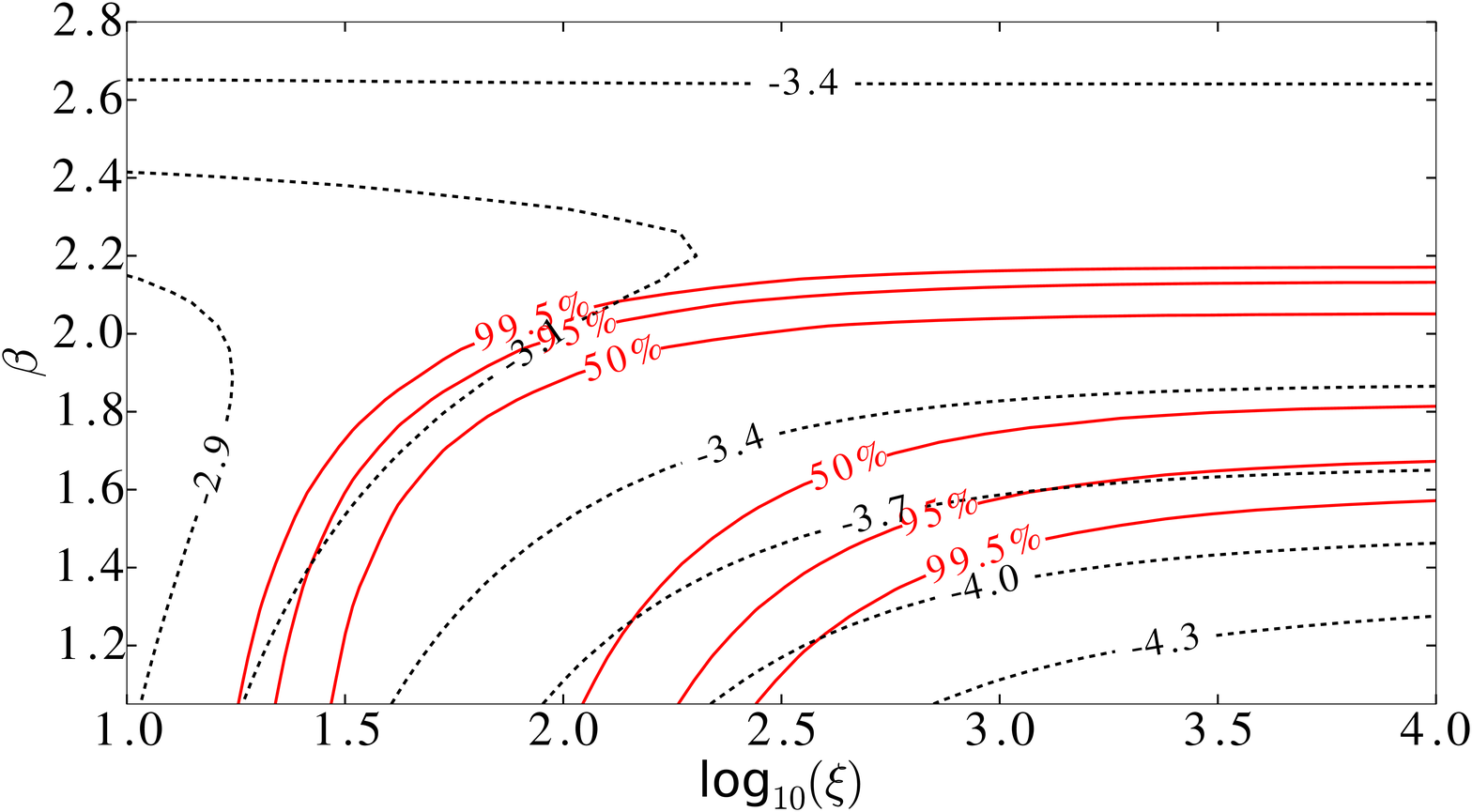}
\caption{The contours of $Q(\beta, \xi)$ defined in
  eq.(\ref{eq:15}). The upper and lower panels are for the 
  two samples with $\flu_{\rm th} = 2$ and $1 \rm\ Jy.ms$
  respectively. Red contours are the confidence levels from the
  $\chi^2$ fitting of each sample.
}\label{fig:Ndet}
\end{figure}

As for the normalization, we draw in fig.(\ref{fig:Ndet}) the
contours of
\begin{equation}
  \label{eq:15}
  Q(\beta, \xi) = \mt{log}_{10}\left[ f_{\rm
      frb}\frac{\tau}{10^3\rm\ d} \frac{\dot{N}_0}{1\rm\ d^{-1}}
    \frac{10^4\rm\ Gpc^3\
    d^{-1}}{\dot{N}_{\rm det}(\flu_{\rm th})}\right]
\end{equation}
for the two samples with $\flu_{\rm th} = 2$ (upper) and $1\rm\
Jy.ms$ (lower panel). Dashed curves are for negative values. For the
parameter space allowed by Parkes' data (red contours,
eq. \ref{eq:16}), we obtain $Q \in (-4.3,-2.7)$. The observational
all-sky FRB rate above $\sim 2\rm\ Jy.ms$ is 
estimated to be $3\times10^3 - 1\times10^4 \rm\ Gpc^{-3}\ d^{-1}$
\citep{2013Sci...341...53T, 2015MNRAS.447.2852K, 2016MNRAS.455.2207R,
  2016MNRAS.tmpL..49C}. Considering all the uncertainties, we obtain
\begin{equation}
  \label{eq:17}
  f_{\rm frb}\frac{\tau}{10^3\rm\ d}  \frac{\dot{N}_0}{1\rm\ d^{-1}}
  \in (10^{-4.8}, 10^{-2.7}).
\end{equation}
However, the normalization parameter $\dot{N}_0$ is still
unconstrained so far and will be discussed in 
next section.

\section{Is FRB 121102 representative?}\label{p0rep}
\begin{figure}
  \centering
\includegraphics[width = 0.45 \textwidth,
  height=0.35\textheight]{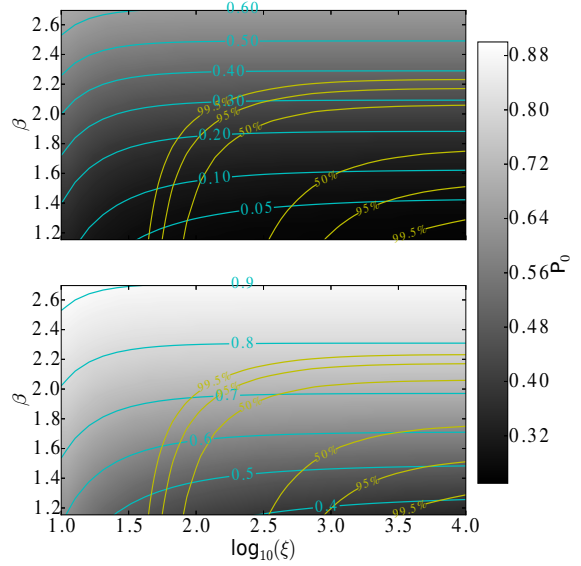}
\caption{The probability of observing none of the 8 bursts repeating
  above the threshold $\flu_{\rm th} = 2\rm\ Jy.ms$. The upper and lower
  panels are for $\dot{N}_0 = 2$ and $0.5\rm\ d^{-1}$
  respectively. Yellow contours are the $\chi^2$ fitting confidence levels from the
  upper panel of fig.(\ref{fig:chisq}). 
}\label{fig:P0rep}
\end{figure}

\citet{2015MNRAS.454..457P} conducted a follow-up survey of the fields
of 8 known FRBs\footnote{They are not the same 8 bursts in
  the sample selected by $\flu_{\rm th} = 2\rm\ Jy.ms$ as in Section
  \ref{sec:obs} (only 3 of them overlap).} and found none
repeating. Other follow-up surveys, 
e.g. \citet{2015ApJ...799L...5R}, are less constraining. For a given
burst $i$ at redshift $z^{\rm i}$, the average number of repeating
events above the 
threshold $\flu_{\rm th}$ within a monitoring time $\Delta T^{\rm i}$
is 
\begin{equation}
  \label{eq:18}
  \overline{N}_{\rm rep}^{\rm i}
  = \frac{\dot{N}_0 \Delta T^{\rm i}/(1+z^{\rm i})}{1-\xi^{1-\beta}}
  \left[\left(\frac{\flu_{\rm th}}{\flu_0}\right)^{1-\beta}  - \xi^{1-\beta}
\right].
\end{equation}
The probability of observing none of the 8 repeating is
\begin{equation}
  \label{eq:19}
  P_0 = \prod_{i=1}^{8} \mt{exp}(-\overline{N}_{\rm rep}^{\rm i}).
\end{equation}
We take the monitoring time $\{\Delta T^{\rm i}\}$ from
\citet{2015MNRAS.454..457P} and assume that the survey is sensitive to
any repeating bursts above the threshold $\flu_{\rm th} = 2\rm\ Jy.ms$. We
calculate $P_0$ as a function of $\beta$ and $\xi$ from
eq.(\ref{eq:18}) and (\ref{eq:19}) 
and show the result in fig.(\ref{fig:P0rep}) for $\dot{N}_0 = 2$
(upper) and $0.5\rm\ d^{-1}$ (lower pannel). 
We find that, if all FRBs repeat at the same rate as FRB 121102
($\dot{N}_0 = 2\rm\ d^{-1}$), the probability of observing none
repeating is $P_0\simeq 5\%$ -- $30\%$. This, again, supports FRB
121102 as being representative of the whole population. Note that
\citet{2015MNRAS.454..457P} used a relatively short cutoff of
de-dispersed width $W\lesssim 8\rm\ ms$, which could lead to
incompleteness above $2\rm\ Jy.ms$ and further increases $P_0$.

On the other hand, \citet{2015MNRAS.447..246P} detected a new FRB at
redshift $z\simeq 0.44$ in the survey. The average number of bNSs
at redshift $\leq z$ in 
a sky area of $8\times0.5 = 4\rm\ deg^2$ is
\begin{equation}
  \label{eq:12}
  \overline{N}(z) =  \frac{4\ \mt{deg^2}\ f_{\rm frb}\tau}{4.1\times10^4\rm\
    deg^2} \int_0^z\md z^\prime \Phi_{\rm 
    cc}(z^\prime) (1+z^\prime) \frac{\md V}{\md z^\prime}.
\end{equation}
The probability of having at least one bNS with
$z\leq 0.44$ in the observed area is
\begin{equation}
  \label{eq:22}
  P_{\rm \geq 1} = 1 - \mt{exp}(-\overline{N}(0.44)) \simeq 1 - \mt{exp}(-1.6
  f_{\rm frb}\tau).
\end{equation}
The product $f_{\rm frb}\tau \dot{N}_0$ is constrained by
eq.(\ref{eq:17}), so we have
\begin{equation}
  \label{eq:23}
\mt{exp}(-\frac{3.2\rm\ d^{-1}}{\dot{N}_0})  < 1- P_{\rm \geq 1} <
\mt{exp}(-\frac{2.5\times 10^{-2}\rm\ 
  d^{-1}}{\dot{N}_0}) .
\end{equation}
If every FRB repeats similarly as FRB 121102 ($\dot{N}_0 = 2\rm\
d^{-1}$), the probability of having $\geq 1$ FRBs  at $z\leq 0.44$ in a
4 $\rm deg^2$ area is $1.2\% < 
P_{\geq 1}<80\%$, with lower $\dot{N}_0$ giving larger $P_{\geq 1}$.

Putting $P_0$ and $P_{\geq 1}$ together, we find the
observations by \citet{2015MNRAS.454..457P} favor a normalization
constant $\dot{N}_0$ smaller than the value of FRB 121102, but
$\dot{N}_0 = 2\rm\ d^{-1}$ is not ruled out at a high confidence. A
relatively steep power-law with $\beta\in (1.5, 2.2)$ is also favored,
as can be seen in fig.(\ref{fig:P0rep}). Therefore, we conclude that FRB
121102 may be slightly more active than average (with a relatively
large $\dot{N}_0$) but is so far consistent with being representative
of the ensemble. 

\section{Summary and Implications for the FRB
  progenitors}\label{sec:summary}
This {\it Letter} is based on three hypotheses: (I) FRBs are produced
by NSs at cosmological distances with their DMs mostly due to free electrons
in the IGM; (II) FRB rate tracks core-collapse
supernova rate; (III) FRBs repeat with a universal EDF $\md \dot{N}/\md
E \propto E^{-\beta}$ with a high-end cutoff 
at $E_{\rm max}$. Based on these assumptions, we find that observations
so far are consistent with a universal EDF with power-law index
$1.5 \lesssim \beta \lesssim 2.2$, high-end cutoff $\xi\equiv E_{\rm
  max}/E_0 \gtrsim 30$ and normalization $\dot{N}_0 \lesssim 2\rm\
d^{-1}$, where $\dot{N}_0$ is defined as the integrated rate above
$E_0 \simeq 1.2\times 10^{39} f_{\rm r}^{-1} \rm\ erg$ ($f_{\rm r}$ being the
radio emission efficiency). We also put constraints on the
dimensionless product $f_{\rm frb}(\tau/10^3\rm\ d) (\dot{N}_0/1\rm\ d^{-1})\in
(1.6\times10^{-5}, 2.0\times10^{-3})$, where $f_{\rm frb}$ is the
fraction of NSs that are able to produce observable FRBs and $\tau$ is
the length of the bursting phase. Better statistics in the future
could narrow down the uncertainties, if our hypotheses are
correct. Some of the 
implications of our results on the FRB progenitors are as follows:

\begin{itemize}
\item The EDF of FRBs is shallower than the power-law tail of
Crab giant pulses \citep[$2.1 < \beta < 3.5$,][]{2012ApJ...760...64M}
and consistent with magnetar bursts \citep[$\beta\simeq
1.66$,][]{1999ApJ...526L..93G} and other avalanche events explained by
Self-Organized Criticality \citep{1991ApJ...380L..89L,
  2016SSRv..198...47A}.

\item Burst energy can reach $E_{\rm max}\gtrsim 10^{41} f_{\rm
    r}^{-1}\rm\ erg$, which is supported by the
  ``Lorimer'' burst with $E>1.1\times10^{41}f_{\rm r}^{-1}\rm\
  erg$ \citep{2015MNRAS.447.2852K}. 
\item The spacial density of FRB progenitors is
  \begin{equation}
    \label{eq:24}
    f_{\rm frb}\tau \Phi_{\rm cc}(1+z)\in (64,
    8000)(\dot{N}_0/1\rm\ d^{-1})^{-1}
    \rm\ Gpc^{-3},
  \end{equation}
where we have used $\Phi_{\rm cc}(z = 1)$ to get the numerical
values. Although $\dot{N}_0$ still has large uncertainties, the fact
the FRB 121102 is 
consistent with being representative means that $\dot{N}_0$ is likely
not much smaller than $1\rm\ d^{-1}$. If we want to use
bNSs to probe the density fluctuations of IGM
\citep{2014ApJ...780L..33M}, the spacial resolution may be $\gtrsim
50\rm\ Mpc$.
\item The fraction of NSs that are able to produce observable FRBs is
  $f_{\rm frb}\in (1.6\mt{e-}5, 2.0\mt{e-}3) (\tau/10^3\rm\
  d)^{-1} (\dot{N}_0/1\rm\ d^{-1})^{-1}$. If we consider the possibility that
  $\tau\sim \rm\ kyr$, the fraction could be as low as $f_{\rm frb}\sim
  10^{-6}$! Although the true fraction is a factor of $4{\mt
    \pi}/\Omega_{\rm r}$ higher ($\Omega_{\rm r}$ is the
  total sky coverage of all bursts from one 
  progenitor), FRB progenitors may be a rare type of NSs. For example,
  the birth rate of magnetars is $\sim 10\%$ of  
  core-collapse rate \citep{2008MNRAS.391.2009K}, so ``being a
  magnetar'' may not be sufficient for FRBs.
\end{itemize}

\section{acknowledgments}
This research was funded by the Named Continuing
Fellowship at the University of Texas at Austin.

\label{lastpage}
\end{document}